\documentclass[iop]{emulateapj}
\usepackage{float}
\usepackage{color}
\usepackage{booktabs}
\usepackage{amsmath}

\usepackage{graphicx}
\usepackage{bm}
\usepackage{epstopdf}

\def\apj{Astrophys. J.}
\def\prd{Phys. Rev. D}
\def\mnras{Mon. Not. R. Astron. Soc.}
\def\jcap{J. Cos. Astropart. Phys.}
\def\aap{Astron. Astrophys.}
\def\araa{Ann. Rev. Astron. Astrophys.}

\def\pasp{Publications of the Astronomical Society of the Pacific}
\def\apjl{Astrophys. J. Letter}
\def\pasa{Publ. Astron. Soc. Aust.}

\begin{document}
\title{Neutrinos from choked jets accompanied by type-II supernovae}
\author{Hao-Ning He$\!$\altaffilmark{1,2,3},
Alexander Kusenko$\!$\altaffilmark{3,4},
Shigehiro Nagataki$\!$\altaffilmark{2,5,6},
Yi-Zhong Fan$\!$\altaffilmark{1},
Da-Ming Wei$\!$\altaffilmark{1}
}

\altaffiltext{1}{Key Laboratory of Dark Matter and Space Astronomy, Purple Mountain Observatory, Chinese Academy of Sciences, Nanjing 210008, China}
\altaffiltext{2}{Astrophysical Big Bang Laboratory, RIKEN, Wako, Saitama, Japan}
\altaffiltext{3}{Department of Physics and Astronomy, University of California, Los Angeles, CA 90095-1547, USA}
\altaffiltext{4}{Kavli IPMU (WPI), University of Tokyo, Kashiwa, Chiba 277-8568, Japan}
\altaffiltext{5}{Interdisciplinary Theoretical Science Research Group (iTHES), RIKEN,
Saitama 351-0198, Japan}
\altaffiltext{6}{Interdisciplinary Theoretical $\&$ Mathematical Science Program (iTHEMS), 
RIKEN, Saitama 351-0198, Japan}

\begin{abstract}
The origin of the IceCube neutrinos is still an open question.
Upper limits from diffuse gamma-ray observations suggest that the neutrino sources are either distant or hidden from gamma-ray observations.  It is possible that the neutrinos are produced in jets that are formed in the core-collapsing massive stars and fail to break out, the so-called choked jets.  We study neutrinos from the jets choked in the hydrogen envelopes of red supergiant stars. 
Fast photo-meson cooling softens the neutrino spectrum, making it difficult to explain the PeV neutrinos observed by IceCube in a one-component scenario, but a two-component model can explain the spectrum. Furthermore, we predict that a newly born jet-driven type-II supernova 
may be observed to be associated with a neutrino burst detected by IceCube. 
\end{abstract}

\maketitle

\section{Introduction}
The origin of high-energy neutrinos observed
by the IceCube observatory is still under debate. 
The distribution of the observed neutrino events is consistent with being isotropic, 
suggesting extragalactic sources~~\citep{IceCube2014ThreeYears}.
Active galactic nuclei~\citep[AGNs;][]{Stecker1991AGN,Essey2010PRL,Essey2011ApJ,Kalashev2013PRL,Padovani2014,Kimura2015AGN,Murase2016HiddenAccelerators}, gamma-ray bursts
~\citep[GRBs;][]{Waxman1997GRB,Murase2013ChokedGRB,Liu2013GRB,Cholis2013GRB,Murase2013Hadronuclear}, 
starburst/star-forming galaxies~\citep{Loeb2006SBG,He2013ULIRG,Tamborra2014SFG,Chang2014SBG,Liu2014HNRinSFG}, 
supernova remnants~\citep{Mandelartz2015SNR,Xiao2016WDM,Zirakashvili2016TypeIInSN} and 
young pulsars~\citep{Murase2009Magnetar,Fang2014pulsar} are among possible candidates.

The flavor composition of the neutrino flux is consistent with the standard scenarios where neutrinos are produced via
pp collision or photo-meson interaction, 
but disfavor a neutron-decay scenario at 3.6$\sigma$ significance~\citep{IceCube2015Combined}.

In addition to the spatial distribution of neutrino events and
the flavor composition, the features of the observed neutrino spectrum are also crucial for identifying the sources.
The IceCube Collaboration has reported a neutrino spectrum that can be best fit by a soft unbroken power-law spectrum with index of $-2.50\pm0.09$ for neutrinos with energies above $10$ TeV. A single power-law neutrino spectrum with index $-2$ is disfavored at the level of  $3.8\sigma$~\citep{IceCube2015Combined}.
An even softer spectrum with index of $-2.92^{+0.33}_{-0.29}$ from six years of High Energy Starting Event (HESE) IceCube data at energies above $40$~TeV was reported~\citep{IceCube2017ICRCII}.
However, cosmic muon neutrinos with energies $>194$ TeV
from the northern hemisphere from six years of IceCube data
have a hard spectral index of $-2.13\pm 0.13$~\citep{IceCube2016MuonNeutrinos}.
The difference between the best-fit spectral indices in different energy bands can be explained by a two-component model\citep{He2013}: a soft component explaining the lower-energy neutrinos,
and a hard component explaining the higher-energy neutrinos. 
This model can be tested with improved statistics.

The flux of the observed all-flavor neutrinos around 
$30~{\rm TeV}$ is as high as $10^{-7}~{\rm GeV cm^{-2} s^{-1} sr^{-1}}$~\citep{IceCube2015ICRCII}.
Meanwhile, the {\it Fermi} Large Area Telescope (LAT) collaboration reported that
blazars' contributions dominate the diffuse extragalactic gamma-ray background (EGB) above $50$ GeV~\citep{Ackermann201650GeV},
while blazars only contribute to less than 10$\%$ of the neutrinos observed by the IceCube.
This leads to the puzzling fact
that diffuse gamma-ray emission from neutrino sources 
is much weaker than one would expect based on IceCube neutrino observations~\citep{Murase2013Hadronuclear,Murase2016HiddenAccelerators}. 
The tension can be avoided if the neutrino sources are distant~\citep{Chang2016Distance}, 
if the neutrino production occurs in $p\gamma$ interactions along the line of sight~\citep{Essey2010PRL,Essey2011ApJ,Kalashev2013PRL}, 
or if the gamma-ray emission in the GeV--TeV band is suppressed, 
so that the neutrino sources are hidden from gamma-ray observations. 
The hidden sources can be AGN cores~\citep{Stecker2005, Murase2016HiddenAccelerators}, 
choked jets in tidal disruption events of supermassive black holes~\citep{Wang2016TDE}
and a pair of choked jets in core-collapse massive stars~\citep{Meszaros2001ChokedGRB,Razzaque2004SN,Murase2013ChokedGRB,Xiao2014,Senno2016ChokedGRB}.

At the end of its life, a massive star can collapse into a neutron star, a quark star, or a black hole, 
and it can produce an energetic jet.
The core collapse of massive stars which lost their outer layer of hydrogen and helium 
due to strong winds or mass transfer to a companion~\citep{Filippenko2005} 
may produce Type Ib/c supernovae (SNe), 
while the core collapse of massive stars which did not lose their stellar envelope
may produce Type II SNe.
There is no essential difference between the Type Ib/c and II SNe,
though their spectra are superficially different due to the different properties of 
their stellar envelopes. 
For massive stars with strong winds, which blow out most of the materials of their stellar envelope, 
i.e., Wolf-Rayet stars,
the jets with large injected energy can easily break through the star and produce gamma-ray emission, 
usually observed as a GRB, 
which is suggested to be associated with a Type Ib/c SN~\citep{Hjorth2012}.
These GRB jets are believed to accelerate protons  
and produce high-energy neutrinos in interactions of protons with gamma rays~\citep{Waxman1997GRB}.
However, the flux of the diffuse neutrinos from GRBs is not high enough to explain the observed neutrino flux,
and no associations between the observed GRBs and neutrinos have been observed so far~\citep{IceCube2015GRB,IceCube2017GRB}.
If the jets are propagating in a thick stellar envelope
or extended material~\citep{Meszaros2001ChokedGRB,Razzaque2004SN,Murase2013ChokedGRB,Xiao2014,Senno2016ChokedGRB}, 
they may not be able to break out through them.
The choked jets might be more ubiquitous than the break-out ones.  
In that case, neutrinos and gamma-rays are produced 
via the interaction of accelerated protons and thermal photons in the choked jets.
Since the neutrinos and gamma rays are produced inside the stellar envelope, 
the source is opaque to gamma-ray photons but transparent for neutrinos.
This can explain the lack of association between the observed GRBs and IceCube neutrinos, 
as well as the tension between the diffuse gamma-ray and neutrino observations.
The detailed features of the observed neutrino spectrum can help us to study the sources hidden in the gamma-ray band,
and the possible association between neutrinos and SNe 
will provide us with more information on the progenitor stars. 

There have been many investigations of choked jets from the progenitor of Type Ib/c SNe,
corresponding to the scenario of failed GRBs,
with the duration of the central engine less than $100$ s,
which is typical for long GRBs
\citep{Meszaros2001ChokedGRB,Razzaque2004SN,Murase2013ChokedGRB,Senno2016ChokedGRB}.
However, for jets choked in the stellar envelope of red/blue supergiant stars, associated with Type II SNe,
the duration of the central engine could be longer than for long GRBs~\citep{Xiao2014};
the feasibility of the long duration is discussed in Section 2.
With the same energy budget,
the scenario with longer duration and lower luminosity 
will lead to a neutrino spectrum with a lower cutoff energy at the high energy end, 
as will be discussed in Section 5.

Since the rate of Type II SNe is about three times that of Type Ib/c SNe~\citep{Li2011}, 
the study of the choked jets from the progenitors of Type II SNe
is also important.
In this paper, we study neutrinos associated with Type II SNe, 
i.e., neutrinos from jets choked in the hydrogen envelope of red supergiant stars (RSGs).
We discuss the conditions under which jets are choked and protons are accelerated in Sections \ref{dynamics} and \ref{acceleration}, respectively.
Then we discuss the interactions between protons and the target photons in Section \ref{targetphotons},
and study the features of the produced neutrinos spectrum in Sections \ref{features} - \ref{Diffuse}.
We predict that a newly born jet-driven type II SN may be observed to be associated with the neutrino bursts in Section \ref{typeIISN}. 
Finally, we discuss and summarize in Section \ref{discussions}.

\section{choked jet dynamics} \label{dynamics}
It is widely believed that, at the end of their lives, a fraction of rapidly rotating massive stars will undergo core collapse, form a compact star or a black hole, and launch a pair of jets.
We assume that the core of an RSG is surrounded by a helium envelope of size $\sim 10^{11}~{\rm cm}$, 
and a hydrogen envelope of size $\sim3\times 10^{13}$ cm
with a slowly varying density $\rho_{\rm H}\simeq 10^{-7}~\rm g cm^{-3}~\rho_{\rm H,-7}$~\citep{Meszaros2001Jet}.
A successful jet will propagate in the helium and hydrogen envelopes, 
and finally break out through the star.
However, if the jet lifetime is shorter than the jet crossing time, 
the jet is stalled before it breaks through the star~\citep{MacFadyen2001, Meszaros2001Jet, Meszaros2001ChokedGRB}.

A forward shock and a reverse shock are produced when the jet is propagating in the hydrogen envelope.
The shocked region between the two shocks is the jet head. 
The Lorentz factor of the jet head and the un-shocked jet plasma is $\Gamma_{\rm h}$ and $\Gamma$,
and the Lorentz factor of the shocked jet plasma
in the rest frame of the un-shocked jet plasma is 
\begin{equation}
\bar{\Gamma}=\Gamma\Gamma_{\rm h}(1-\beta\beta_{\rm h}),
\end{equation}
where $\beta=\sqrt[]{1-1/\Gamma}$ and $\beta_{\rm h}=\sqrt[]{1-1/\Gamma_{\rm h}}$.
The energy density of the region behind the reverse shock is ~\citep{Blandford1976,Sari1995}
\begin{equation}
e_{\rm r}=(4\bar\Gamma+3)(\bar\Gamma-1)n_{\rm j}m_{\rm p}c^2,
\end{equation}
where $n_{\rm j}=L/(4\pi \Gamma^2R_{\rm h}^2 c m_{\rm p} c^2)$ is the number density of the un-shocked jet,
with the radius $R_{\rm h}\simeq 2\beta_{\rm h} c t$.
The energy density of the region behind the forward shock is
\begin{equation}
e_{\rm f}=(4\Gamma_{\rm h}+3)(\Gamma_{\rm h}-1)\rho_{\rm H}c^2.
\end{equation}
Balancing the energy densities of these two regions,
one can derive the Lorentz factor and the radius of the jet head, $\Gamma_{\rm h}$ and $R_{\rm h}$.

According to Equation (1) in \citet{Meszaros2001Jet},
for a larger isotropic luminosity, i.e.,  $L_{\rm iso}>10^{54}~{\rm erg~ s^{-1}}~r_{13.5}^{2}\rho_{{\rm H},-7}$,
the velocity of the jet head is relativistic ~\citep{Razzaque2004SN, Ando2005SN, Murase2013ChokedGRB,Senno2016ChokedGRB}.
In this paper, we scan the parameter space of the isotropic luminosity $L_{\rm iso}$ and the lifetime $t$ of the jet,
and find that, for most of the parameters we are concerned with, the jet head is non-relativistic, i.e., $\Gamma_{\rm h}\sim1$,
allowing us to perform an analytical calculation as follows.

In the non-relativistic case, the velocity of the jet head at radius $r$ is approximated to be
\begin{equation}\label{v}
v_{\rm h}\simeq\left(\frac{L_{\rm iso}}{(4\Gamma_{\rm h}+3)\pi r^2 c\rho_{\rm H}}\right)^{1/2}
\end{equation}
and the corresponding propagating time is 
\begin{equation}\label{t}
t_{\rm prop}=\int\frac{dr}{v_{\rm h}}\simeq \frac{R_{\rm h}}{2v_{\rm h}}.
\end{equation}
Combining Equations (\ref{v}) and (\ref{t}),
one can derive the radius and the velocity of the jet head at the end of the jet lifetime $t$ analytically,
i.e.,
\begin{equation}\label{R_h}
R_{\rm h}=9.0\times10^{12}~{\rm cm}~L_{\rm iso,48}^{1/4}t_4^{1/2}\rho_{{\rm H},-7}^{-1/4},
\end{equation}
and 
\begin{equation}
v_{\rm h}=0.015~c~L_{\rm iso,48}^{1/4}t_4^{-1/2}\rho_{{\rm H},-7}^{-1/4}.
\end{equation}
The above results are consistent with the Lorentz factor of the jet head being $\Gamma_{h}\simeq1$
as long as 
\begin{equation}\label{non-relativistic}
L_{\rm iso,48}t_4^{-2}\leqslant 1.2\times10^6~\rho_{{\rm H},-7}.
\end{equation}
According to Equation (\ref{R_h}), the jet crossing time, i.e., the time that the jet takes to break through the stellar envelope of radius of $R$, is 
\begin{equation}
t_{\rm cros}=1.1\times10^{5}~{\rm s}~R_{13.5}^2L_{\rm iso,48}^{-1/2}\rho_{{\rm H},-7}^{1/2}
\end{equation}
If the jet lifetime is smaller than the jet crossing time, i.e.,
\begin{equation}\label{chokedjetconstraint}
t<t_{\rm cros},
\end{equation} 
the jet cannot break out through the stellar envelope.
In this case, the jet is choked and this object will not be observed as a shock breakout phenomenon or a GRB. 
Rather, it will become a jet-driven type II SN. We will discuss this point in Section \ref{typeIISN}.

According to observations, the duration of long GRBs is about $<100~{\rm s}$, 
while very long GRBs and ultra-long GRBs with durations of $\sim10^3~{\rm s}$
and $\sim10^4~{\rm s}$ have been observed~\citep{Greiner2015},
and a few GRBs can reach $\sim 10^5~{\rm s}$,
with the central engine unclear~\citep{Levan2015}.
The majority of the durations of the choked jet events are plausibly
different from those of the successful jet events, i.e., the observed GRBs.
If we assume the jet is powered by the rotation energy of the central magnetar~\citep{Usov1992,Metzger2011,Mazzali2014,Greiner2015,Gompertz2017} 
with a mass of $1.4~M_\odot$ and a radius of $10~{\rm km}$,
the upper limit of the jet lifetime is about the spin-down duration of the magnetar,
which is~\citep{Ostriker1969} 
\begin{equation}
t_{\rm sd}=2.0\times 10^5~{\rm s}~P_{\rm i,-3}^2B_{\rm m,14}^{-2},
\end{equation}
where $P_{\rm i}$ is the initial spin period,
and $B_{\rm m}$ is the strength of magnetic field of the magnetar.
If we assume the jet is powered by the accretion of the star material onto the central black hole,
the upper limit of the jet lifetime is the free-fall time of the star material,
which is~\citep{Kippenhahn1994}
\begin{equation}
t_{\rm fb}=\left(\frac{\pi^2R^3}{8GM_{\rm c}}\right)^{\frac{1}{2}}=1.7\times10^7~{\rm s}~R_{13.5}^{3/2}
\left(\frac{M_{\rm c}}{M_{\odot}}\right)^{-\frac{1}{2}},
\end{equation}
where $G$ is the gravitational constant, and $M_{\rm c}$ is the mass of the central black hole.
Therefore, in this paper, we assume the jet lifetime to be a free parameter
with a value in the range of $\sim10~-~10^6~{\rm s}$.

\section{Particle Acceleration Constraint in the internal shock}\label{acceleration}
Due to the inhomogeneities of the jet,
an internal shock can be produced beneath the jet head at radius $R_{\rm IS}\lesssim R_{\rm h}$, 
by a rapid shell catching up with and merging into the slow shell.
The relative Lorentz factor between the two shells is 
$\Gamma_{\rm rel}\simeq(\Gamma_{\rm r}/\Gamma+\Gamma/\Gamma_{\rm r})/2\sim 3$,
where $\Gamma_{\rm r}$ and $\Gamma$ are the Lorentz factors of the rapid and merged shells, respectively.
Cosmic rays are accelerated efficiently in the internal shock, 
if the comoving size of the upstream flow $l_{\rm u}=\frac{R_{\rm IS}/\Gamma}{\Gamma_{\rm rel}}$ is much smaller than
the mean free path length of the thermal photons in the upstream flow, 
$l_{\rm dec}=1/(n_{\rm u}\sigma_{\rm T})$ ~\citep{Murase2013ChokedGRB},
where the number density of particles in the upstream flow is
$n_{\rm u}=\frac{n_{\rm p,IS}}{\Gamma_{\rm rel}}$, and 
$n_{\rm p,IS}=\frac{L_{\rm iso}}{4\pi R_{\rm IS}^2\Gamma^2m_{\rm p}c^3}=2.2\times10^{11}~{\rm cm^{-3}}~\Gamma_{1}^{-2}L_{\rm iso,48}^{1/2}t_{4}^{-1}\rho_{\rm H,-7}^{1/2}$ is the proton number density in the un-shocked material.
This acceleration constraint of $l_{\rm u}<l_{\rm dec}$ requires that~\citep{Murase2013ChokedGRB, Senno2016ChokedGRB},
\begin{equation}
\tau=n_{\rm p,IS}\sigma_{\rm T}(R_{\rm IS}/\Gamma)<{\rm min}[\Gamma_{\rm rel}^{2},~0.1C^{-1}\Gamma_{\rm rel}^3],
\end{equation}
with 
$C=1+2ln\Gamma_{\rm rel}^2$ is a coefficient accounting for pair production.
When the internal shock approaches the jet head, i.e., $R_{\rm IS}\sim R_{\rm h}$,
the criterion of efficient acceleration of protons can be rewritten as
\begin{equation}
\begin{split}
\tau&=0.13~\Gamma_{1}^{-3}L_{\rm iso,48}^{3/4}t_{4}^{-1/2}\rho_{\rm H,-7}^{1/4}\\
&<{\rm min}[\Gamma_{\rm rel}^{2},0.1C^{-1}\Gamma_{\rm rel}^3],
\end{split}
\label{radiationconstraint}
\end{equation}
Then the accelerated protons interact with the photons escaping from the jet head.

If the constraint shown in Equation (\ref{radiationconstraint}) is satisfied,
the internal shock is able to accelerate protons,
with an acceleration timescale of 
\begin{equation}\label{t_acc}
t_{\rm p, acc}(\epsilon_{\rm p})=\phi\frac{\epsilon_{\rm p}}{q_{\rm e} Bc}\\
=0.14~{\rm s}~\phi_{1}\epsilon_{B,-2}^{-1/2}\epsilon_{{\rm p},15}L_{\rm iso,48}^{-1/4}\Gamma_{1}t_{4}^{1/2}\rho_{{\rm H},-7}^{-1/4},
\end{equation}
where 
$\phi$ is the number of the gyro-radii required to {\it e}-fold the particle energy,
$q_{\rm e}$ and $\epsilon_{\rm p}$ are the charge and the energy of the proton,
and 
\begin{equation}
B=9.1\times10^3~{\rm G}~\epsilon_{B,-2}^{1/2}\Gamma_{1}^{-1}L_{\rm iso,48}^{1/4}t_4^{-1/2}\rho_{{\rm H},-7}^{1/4},
\end{equation}
is the strength of the magnetic field in the internal shock 
with $\epsilon_{\rm B}$ as the fraction of the magnetic field energy.

\section{the target photons}\label{targetphotons}
In the shocked jet head region, the Thomson optical depth of the shocked plasma is
\begin{equation}
\begin{split}
\tau_{\rm h}&=n_{\rm h}\sigma_{\rm T}R_{\rm h}\\
&=52~\Gamma_{1}^{-1}L_{\rm iso,48}^{3/4}t_{4}^{-1/2}\rho_{\rm H,-7}^{1/4}f_{\rm a},
\end{split}
\end{equation}
where $n_{\rm h}=(4\bar\Gamma+3)n_{\rm j}$ is the number density of electrons in the shocked
jet head region, and $f_{\rm a}=\frac{4\bar\Gamma+3}{4\Gamma}\sim 1$ if $\Gamma\gg1$ and $\Gamma_{\rm h}\simeq 1$.
If 
\begin{equation}\label{tau_h}
\tau_{\rm h}>1,
\end{equation}
the electrons in the reverse shock are heated and lose all their energy $\epsilon_{\rm e}e_{\rm r}$
into a thermal radiation with a temperature of
\begin{equation} 
k_{\rm B}T_{\rm h}=99~{\rm eV}~{\epsilon}_{\rm e,-1}^{1/4}L_{\rm iso, 48}^{1/8}t_{4}^{-1/4}
\rho_{{\rm H},-7}^{1/8}f_{\rm c},
\end{equation}
where $k_{\rm B}$ is the Boltzmann constant, and $f_{\rm c}=\left(\frac{(4\bar\Gamma+3)(\bar\Gamma-1)}{4\Gamma^2}\right)^{1/4}\sim1$ if $\Gamma\gg 1$ and $\Gamma_{\rm h}\simeq 1$.
The number density of thermal photons in the shocked jet is
\begin{equation}
\begin{split}
n_{\gamma, \rm h}&=19\frac{\pi}{{(hc)}^3}(k_{\rm B}T_{\rm h})^3\\
&=3.1\times10^{19}~{\rm cm^{-3}}~{\epsilon}_{\rm e,-1}^{3/4}L_{\rm iso, 48}^{3/8}t_{4}^{-3/4}
\rho_{{\rm H},-7}^{3/8}f_{\rm c}^3
\end{split}
\end{equation}

When the internal shock approaches the jet head region, 
the thermal photons will escape into the internal shock with a fraction of $f_{\rm esc}=1/\tau_{\rm h}$~\citep{Murase2013ChokedGRB},
since it's optically thin in the internal shock. 
There, the energy of the thermal photons from the jet head is boosted by a factor of 
$\bar\Gamma\sim \Gamma$ if $\Gamma\gg\Gamma_{\rm h}$ and $\Gamma_{\rm h}\simeq1$.
Then the number density of thermal photons in the internal shock 
is 
\begin{equation}
\begin{split}
n_{\gamma,{\rm IS}}&=\bar{\Gamma}n_{\gamma,{\rm h}}f_{\rm esc}\\
&=6.0\times10^{18}~{\rm cm^{-3}}~\epsilon_{\rm e,-1}^{3/4}\Gamma_{1}^2L_{\rm iso, 48}^{-3/8}t_{4}^{-1/4}
\rho_{{\rm H},-7}^{1/8}f_{\rm c}^3f_{\rm a}^{-1}
\end{split}
\end{equation}
and the peak energy of the thermal photons in the internal shock frame is boosted to be
\begin{equation}\label{EgIS}
\epsilon_{\gamma,\rm IS}=\bar\Gamma(2.8k_{\rm B}T_{\rm h})
=2.8\times10^3~{\rm eV}~\epsilon_{\rm e,-1}^{1/4}\Gamma_{1}L_{\rm iso, 48}^{1/8}t_{4}^{-1/4}\rho_{{\rm H},-7}^{1/8}f_{\rm c}.
\end{equation}

On the other hand,
the electrons are accelerated in the internal shock.
The acceleration timescale for electrons with Lorentz factor of $\gamma_{\rm e}$
is 
\begin{equation}
\begin{split}
t_{\rm e,acc}(\gamma_{\rm e})&=\phi_{\rm e}\frac{\gamma_{\rm e}m_{\rm e}c^2}{q_{\rm e}Bc}\\
&=2.1\times10^{-9}~{\rm s}~\phi_{\rm e,1}\gamma_{\rm e,1.5}\epsilon_{B,-2}^{-1/2}\Gamma_{1}L_{\rm iso,48}^{-1/4}t_4^{1/2}\rho_{{\rm H},-7}^{-1/4},
\end{split}
\end{equation}\label{te_acc}
and the cooling timescale via synchrotron emission~\citep{Sari1998} is
\begin{equation}
\begin{split}
t_{\rm e,c}(\gamma_{\rm e})&=\frac{\gamma_{\rm e}m_{\rm e}c^2}{\frac{4}{3}\sigma_{\rm T}c\gamma_{\rm e}^2\frac{B^2}{8\pi}}\\
&=0.28~{\rm s}~\gamma_{\rm e,1.5}^{-1}\epsilon_{B,-2}^{-1}\Gamma_{1}^{2}L_{\rm iso,48}^{-1/2}t_4\rho_{{\rm H},-7}^{-1/2},
\end{split}
\end{equation}
where $\phi_{\rm e}$ is the number of the gyro-radii required to {\it e}-fold the electron energy.
Comparing the acceleration timescale with the cooling timescale above,
one can derive the maximum Lorentz factor of electrons accelerated in the internal shock,
which is
\begin{equation}
\gamma_{\rm e,max}=3.8\times10^5~\phi_{\rm e,1}^{-1/2}\epsilon_{B,-2}^{-1/4}\Gamma_{1}^{1/2}L_{\rm iso,48}^{-1/8}t_4^{1/4}\rho_{{\rm H},-7}^{-1/8}
\end{equation}
We assume a fraction of $\epsilon'_{\rm e}$ of the shock energy goes into 
the electrons, and the spectral index of the accelerated electrons is $p~=~2.2$,
then the minimum Lorentz factor of the accelerated electrons in the internal shock is~\citep{Sari1998}
\begin{equation}
\gamma_{\rm e,min}=\epsilon'_{\rm e}\frac{(2-p)}{(1-p)}\frac{m_{\rm p}}{m_{\rm e}}=31~\epsilon'_{\rm e,-1}.
\end{equation}
So the cooling timescale for electrons with Lorentz factor $\gamma_{\rm e,min}$ is $t_{\rm e,c}(\gamma_{\rm e, min})=0.30~{\rm s}~{\epsilon'_{\rm e}}_{\rm ,-1}^{-1}\epsilon_{B,-2}^{-1}\Gamma_{1}^{2}L_{\rm iso,48}^{-1/2}t_4\rho_{{\rm H},-7}^{-1/2}$,
which is much smaller than the dynamic timescale of the shock
\begin{equation}
t_{\rm dyn}=R_{\rm IS}/(\Gamma c)=30~{\rm s}~\Gamma_{1}^{-1}L_{\rm iso,48}^{1/4}t_4^{1/2}\rho_{{\rm H},-7}^{-1/4}.
\end{equation}\label{dynamictimescale}
In addition, from Equation (\ref{te_acc}), one has $t_{\rm e,c}(\gamma_{\rm e})\propto \gamma_{\rm e}^{-1}$.
Therefore, electrons with Lorentz factor $\gamma_{\rm e}>\gamma_{\rm e,min}$ are cooled even faster via synchrotron radiation. 
Finally, most of the electrons are cooled to the low-energy end with Lorentz factor $\gamma_{\rm e}<\gamma_{\rm e,min}$.
The energy of the corresponding photons is smaller than~\citep{Sari1998}
\begin{equation}
\begin{split}
\nu_{\rm e,min}^{\rm syn}&=\gamma_{\rm e,min}^2\frac{q_{\rm e}B}{2\pi m_{
\rm e}c}\\
&=0.10~{\rm eV}~\epsilon_{B,-2}^{1/2}\Gamma_{1}^{-1}L_{\rm iso,48}^{1/4}t_4^{-1/2}\rho_{{\rm H},-7}^{1/4},
\end{split}
\end{equation}
which is too low to reach the threshold of photo-meson interaction and interact with the accelerated protons in the internal shock (the maximum energy of the accelerated protons is calculated in Section 5.2.). 
Therefore, the synchrotron photons of electrons
in the internal shock as the target photons of photo-meson interaction can be ignored.
\footnote{We also estimated the density of photons produced by electrons at the high-energy tail $\gamma_{\rm e,min}<\gamma_{\rm e}<\gamma_{\rm e,max}$, and found out that those $\sim {\rm keV}$ synchrotron photons are not dominant over the thermal photons from jet head.}

\begin{figure}
\includegraphics[width=95mm,angle=0]{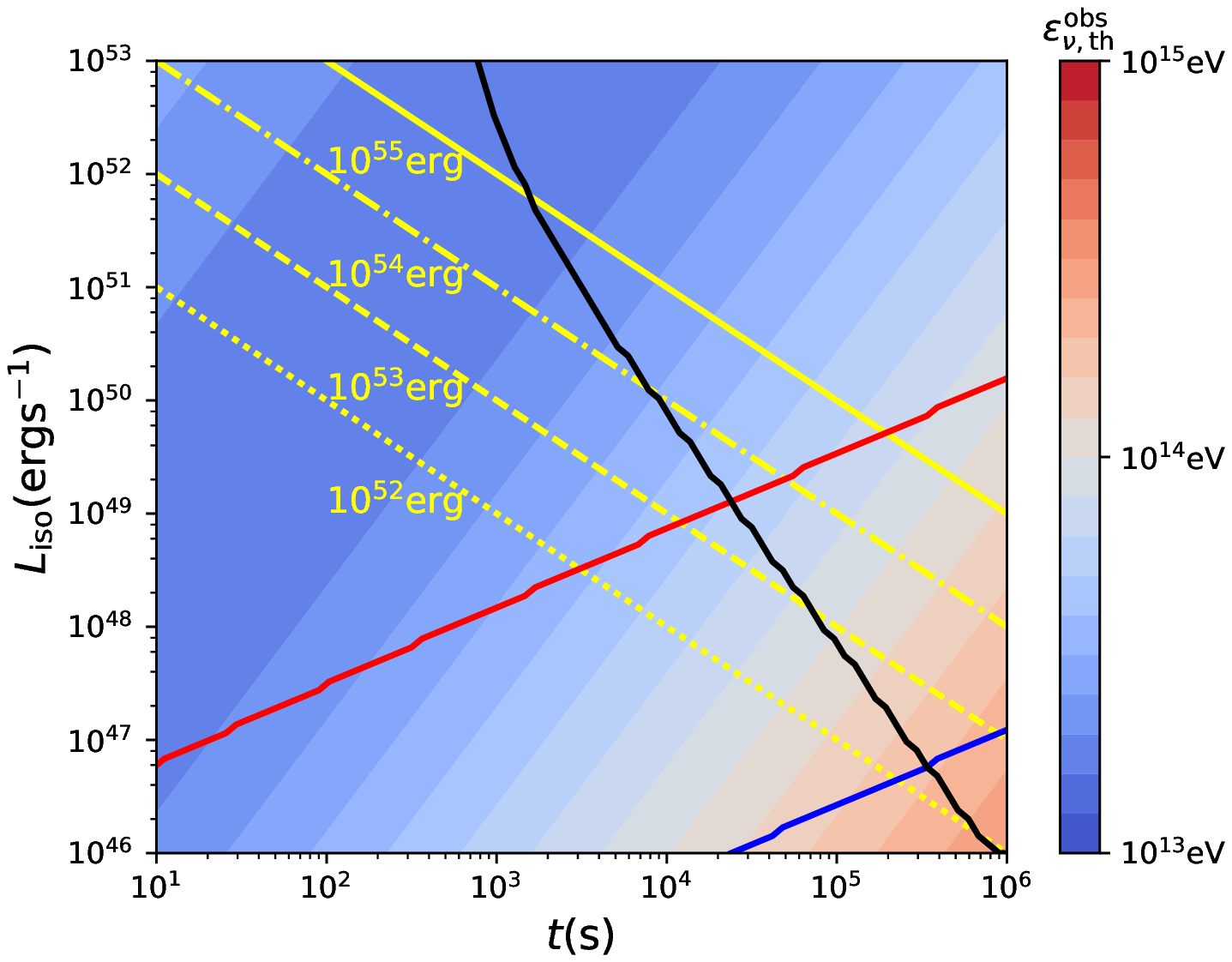}
\includegraphics[width=95mm,angle=0]{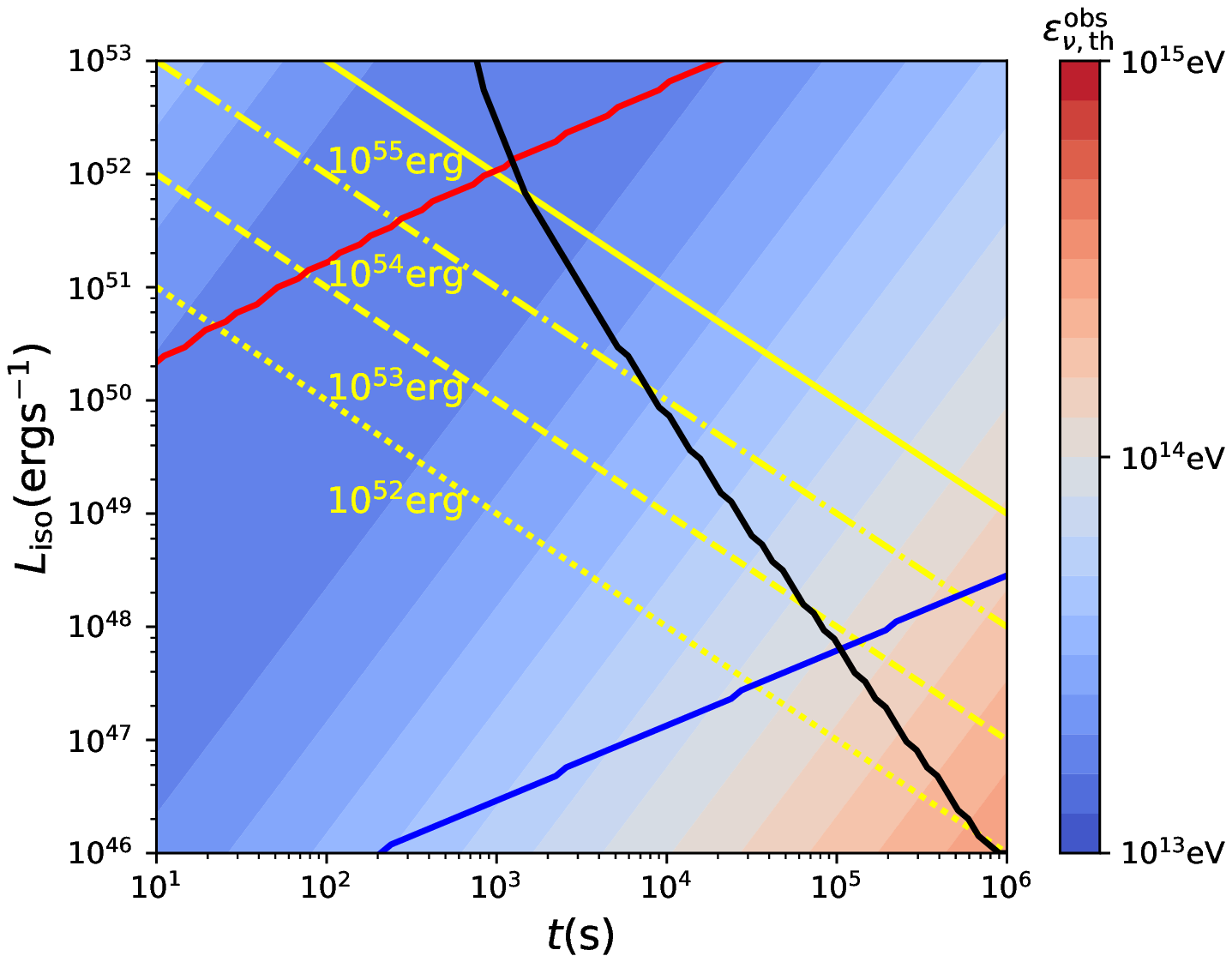}
 \caption{
Constrained luminosity-duration ($L_{\rm iso}-t$) space with fixed Lorentz factor $\Gamma=10$ (upper panel) and $\Gamma=100$ (bottom panel),
respectively.
We fix the outer radius and the density of the hydrogen envelope as $R_{\rm H}=3\times10^{13}~{\rm cm}$ and $\rho_{\rm H}=10^{-7}~{\rm g~cm^{-3}}$,
with $\epsilon_{\rm e}=0.1$ and $\epsilon_{\rm B}=0.01$. 
The parameter space to the left of the black solid line corresponds to the choked jet situation,
that below the red solid line satisfies the acceleration constraint,
and that above the blue solid line corresponds to the jet head optically thick situation.
The color contours denote the values of the low-energy cutoff of the neutrino spectrum, 
which is only valid in the parameter space enclosed by the blue, black, and red solid lines.
The yellow dotted, dashed, dash--dotted and solid lines correspond to the isotropic injected energy of $E_{\rm iso}=10^{52}~{\rm erg},10^{53}~{\rm erg},10^{54}~{\rm erg}$, and $10^{55}~{\rm erg}$, respectively.
}
\label{epsilon_th}
\end{figure} 

\section{features of the neutrino spectrum}\label{features}
As discussed above, our model is valid only if some conditions are met, namely
the jet is choked, protons are accelerated to high energy efficiently, 
and thermal photons are produced in the jet head and propagate into the internal shock region.
Then the accelerated protons interact with the photons from the choked jet head and produce pions. 
A charged pion with the characteristic energy $\epsilon_{\pi}\sim 0.2~\epsilon_{\rm p}$,
decays into a neutrino and a muon with characteristic energy $\epsilon_{\mu}\sim 0.15~\epsilon_{\rm p}$
via the process $\pi^{+}(\pi^{-})\rightarrow \mu^{+}(\mu^{-})+\nu_{\mu}(\bar\nu_{\mu})$.
The produced muon further decays into one lepton and two neutrinos:
$\mu^{+}(\mu^{-})\rightarrow e^{+}(e^{-})+\nu_e(\bar\nu_e)+\bar\nu_\mu(\nu_\mu)$.
The average energy of the neutrinos is $\epsilon_{\nu}\sim 0.05~\epsilon_{\rm p}$.
The produced neutrino spectrum has two features:
one is a cutoff at low energy due to the threshold of the photo-meson interaction,
and the other is a cutoff at high energy due to the photo-meson cooling of protons,
or the synchrotron cooling of pions and muons. Below we calculate the two characteristic 
energies of the neutrino spectrum.

\subsection{The low-energy cutoff }
For a proton energy in the range 
\begin{equation}
0.2~{\rm GeV}^2/(2\epsilon_{\gamma,\rm IS})\leq\epsilon_{\rm p}\leq 0.4~{\rm GeV}^2/(2\epsilon_{\gamma,\rm IS}),
\end{equation}
the $\Delta$-resonance dominates the photo-meson interaction, with cross section $\sigma_\Delta=5\times10^{-28}~{\rm cm}^2$.
Assuming the inelasticity is $f_{\rm in,p\Delta}=0.2$~\citep{Atoyan2001Blazar},
the timescale of the $\Delta$-resonance cooling is approximated to be
\begin{equation}\label{t_Delta}
t_{\Delta}=\frac{1}{\sigma_\Delta n_{\gamma,\rm IS}cf_{\rm in,p\Delta}}\simeq
0.056~{\rm s}~\epsilon_{{\rm e},-1}^{-3/4}L_{\rm iso,48}^{3/8}\Gamma_{1}^{-2}t_{4}^{1/4}
\rho_{{\rm H},-7}^{-1/8}f_{\rm a}f_{\rm c}^{-3},
\end{equation}
which is much smaller than the dynamic timescale of the shock $t_{\rm dyn}$
as in Equation (\ref{dynamictimescale}),
meaning that the photo-meson interaction efficiency is as high as $100\%$.
The timescale of the protons in the internal shock losing energy via the pp collision
is 
\begin{equation}
t_{\rm pp}=\frac{1}{\sigma_{\rm pp}n_{\rm p, IS}cf_{\rm in,pp}}
=1.5\times10^4~{\rm s}~L_{{\rm iso},48}^{-\frac{1}{2}}t_{4}\rho_{{\rm H},-7}^{-\frac{1}{2}}\Gamma_{1}^2,
\end{equation}
where the cross section of the pp collision is approximated to be $\sigma_{\rm pp}\simeq 5\times 10^{-26}~{\rm cm^2}$,
and the inelasticity is assumed to be $f_{\rm in,pp}=0.2$.
The pp collision can be neglected since the timescale of the energy loss $t_{\rm pp}$ is much longer than the dynamic timescale $t_{\rm dyn}$.

For $\epsilon_{\rm p}<0.2~{\rm GeV}^2/(2\epsilon_{\gamma,\rm IS})$,  
the cross section of the photo-meson interaction decreases rapidly,
leading to a cutoff at the low-energy end of the neutrino spectrum~\citep{Mucke1999}.

Therefore, the threshold of the photo-meson interactions in the internal shock frame corresponds to the proton energy 
\begin{equation}
\begin{split}
\epsilon_{\rm p, th}&=0.2~{\rm GeV}^2/(2\epsilon_{\gamma,\rm IS})\\
&=3.6\times10^{13}~{\rm eV}~
\epsilon_{e,-1}^{-1/4}\Gamma_{1}^{-1}L_{\rm iso, 48}^{-1/8}t_{4}^{1/4}
\rho_{{\rm H},-7}^{-1/8}f_{\rm c}^{-1}.
\end{split}
\end{equation}
Thus, a lower-energy cutoff of the observed neutrino spectrum appears due to the threshold of the photo-meson interaction at 
\begin{equation}\label{Enu_th}
\begin{split}
\epsilon_{\nu,{\rm th}}^{\rm obs}&=0.05\epsilon_{\rm p,th}\Gamma\\
&=1.8\times10^{13}~{\rm eV}~\epsilon_{e,-1}^{-1/4}L_{\rm iso,48}^{-1/8}t_{4}^{1/4}\rho_{{\rm H},-7}^{-1/8}f_{\rm c}^{-1}.
\end{split}
\end{equation}

We plot the value of $\epsilon_{\nu,{\rm th}}^{\rm obs}$ as a function of the isotropic luminosity $L_{\rm iso}$ and the lifetime $t$ of the jet
using color contours in Figure \ref{epsilon_th}.
The parameter space to the left of the black solid line satisfies the choking condition as discussed in Section 2.
The parameter space below the red solid curve satisfies the internal shock acceleration constraint as discussed in Section 3.
The parameter space above the blue solid curve satisfies the condition of producing thermal photons in the jet head as discussed in Section 4.
The color contours are valid for parameters which satisfy these constraints, i.e., in the region enclosed by the red, blue, and black solid lines.
Comparing the two panels, the choking condition does not change for different Lorentz factors, 
but the shock acceleration constraint is tighter for a smaller Lorentz factor.
Thus, for a smaller Lorentz factor $\Gamma=10$, 
only jets with a lower luminosity and a longer lifetime can accelerate protons efficiently in the internal shock,
and the low-energy cutoff is at around a few tens to a few hundreds of TeV for local sources. 
For a larger Lorentz factor $\Gamma=100$,
the low energy cutoff of the observed neutrino spectrum is about $1-10$ TeV.

\subsection{The high-energy cutoff}\label{TheHighEnergyCutoff}
For protons with energies  $\epsilon_{\rm p}> 0.4{\rm GeV}^2/(2\epsilon_{\gamma,\rm IS})$,
i.e.,
\begin{equation}
\epsilon_{\rm p}>7.2\times10^{13}~{\rm eV}~
\epsilon_{e,-1}^{-1/4}\Gamma_{1}^{-1}L_{\rm iso, 48}^{-1/8}t_{4}^{1/4}
\rho_{{\rm H},-7}^{-1/8}f_{\rm c}^{-1},
\end{equation}
the other resonances and multi-pion channel will dominate the photo-meson interaction,
with the average cross section approximated to be $\sim1\times10^{-28}~{\rm cm^2}$ ~\citep{Mucke1999}.
Assuming the inelasticity is 0.6~\citep{Atoyan2001Blazar},
the corresponding timescale of multi-pion production is
\begin{equation}\label{tpg}
t_{p\gamma}\simeq
0.093~{\rm s}~\epsilon_{{\rm e},-1}^{-3/4}L_{\rm iso,48}^{3/8}\Gamma_{1}^{-2}t_{4}^{1/4}
\rho_{{\rm H},-7}^{-1/8}f_{\rm a}f_{\rm c}^{-3}.
\end{equation}
When $t_{p\gamma}\geq t_{\rm p,acc}(\epsilon_{\rm p})$,
the protons cannot be accelerated to an energy as high as $\epsilon_{\rm p}$ before cooling. 
\footnote{ Since $t_{\Delta}\ll t_{\rm dyn}$ and $t_{{\rm p}\gamma}\ll t_{\rm dyn}$, the source is considered to be ``calorimetric",
i.e., the protons lose all their energy and cannot escape from the internal shock, thus we do not have to consider the pp collision
between the protons accelerated in the internal shock and hydrogen in the stellar envelope.}
Comparing the photo-meson cooling timescale (Equation (\ref{tpg})) and the acceleration timescale (Equation (\ref{t_acc})),
one can derive the maximum energy of accelerated protons due to the photo-meson cooling
as
\begin{equation}
\begin{split}
\epsilon_{\rm p,c}\simeq & 6.6\times10^{14}~{\rm eV}\\
&\times\phi_{1}^{-1}\epsilon_{{\rm e},-1}^{-3/4}\epsilon_{B,-2}^{1/2}L_{\rm iso,48}^{5/8}\Gamma_{1}^{-3}t_{4}^{-1/4} \rho_{{\rm H},-7}^{1/8}f_{\rm a}f_{\rm c}^{-3}.
\end{split}
\end{equation} 
Consequently, the observed neutrino spectrum has a cutoff at energy 
\begin{equation}\label{Enu_c}
\begin{split}
\epsilon_{\nu,\rm c}^{\rm obs}&=0.05\epsilon_{\rm p,c}\Gamma\sim3.3\times10^{14}~{\rm eV}\\
&\times\phi_{1}^{-1}\epsilon_{{\rm e},-1}^{-3/4}\epsilon_{B,-2}^{1/2}L_{\rm iso,48}^{5/8}\Gamma_{1}^{-2}t_{4}^{-1/4}
\rho_{{\rm H},-7}^{1/8}f_{\rm a}f_{\rm c}^{-3}.
\end{split}
\end{equation}

Moreover, 
the synchrotron cooling of pions and muons will suppress the flux of neutrinos
when the synchrotron cooling timescales of pions and muons are shorter than their lifetimes.
Adopting the approximations that the energy of the produced pions is $\epsilon_{\pi}=0.2\epsilon_{\rm p}$ and
the energy of the produced muons is $\epsilon_{\mu}=0.15\epsilon_{\rm p}$, 
the lifetimes of pions and muons can be written as functions of $\epsilon_{\rm p}$, i.e.,
$\tau_{\pi}=0.038~{\rm s}~\epsilon_{{\rm p},15}$
and 
$\tau_{\mu}=3.15~{\rm s}~\epsilon_{{\rm p},15}$.
The synchrotron cooling timescales of pions and muons are
\begin{equation}
\tau_{\pi,\rm syn}
=1.3\times10^2~{\rm s}~\epsilon_{\rm p,15}^{-1}\epsilon_{B,-2}^{-1}\Gamma_{1}^{2}L_{\rm iso,48}^{-1/2}t_4\rho_{{\rm H},-7}^{-1/2}
\end{equation}
and 
\begin{equation}
\tau_{\mu,\rm syn}
=59~{\rm s}~\epsilon_{\rm p,15}^{-1}\epsilon_{B,-2}^{-1}\Gamma_{1}^{2}L_{\rm iso,48}^{-1/2}t_4\rho_{{\rm H},-7}^{-1/2}
\end{equation}
Comparing the synchrotron cooling timescale with the lifetime of the leptons,
we find that, during the photo-meson interactions, muons with the energy larger than $\epsilon_{\mu,\rm sup}$
cool first before decaying,
where the critical energy of muons in the internal shock is 
\begin{equation}
\epsilon_{\mu,\rm sup}=6.5\times10^{14}~{\rm eV}~\epsilon_{B,-2}^{-1/2}\Gamma_{1}L_{\rm iso,48}^{-1/4}t_4^{1/2}\rho_{{\rm H},-7}^{-1/4}
\end{equation}
Then the observed neutrino spectrum is suppressed at energies above
\begin{equation} \label{Epion_sup}
\epsilon_{\nu,\rm sup}^{\rm obs}\simeq 2.2\times10^{15}~{\rm eV}~\epsilon_{B,-2}^{-1/2}\Gamma_{1}^2L_{\rm iso,48}^{-1/4}t_4^{1/2}\rho_{{\rm H},-7}^{-1/4}
\end{equation}
due to the synchrotron cooling of the muons.

Muons with energy larger than $\epsilon_{\mu,\rm sup}$ will produce
synchrotron photons with the energy larger than the characteristic energy of
\begin{equation}
\begin{split}
E_{\mu,\rm syn}&=\gamma_{\mu,\rm sup}^2\frac{q_{\rm e}B}{2\pi m_{\mu}c}\\
&=1.9\times10^7~{\rm eV}~\epsilon_{B,-2}^{-1/2}\Gamma_{1}L_{\rm iso,48}^{-1/4}t_4^{1/2}\rho_{{\rm H},-7}^{-1/4}
\end{split}
\end{equation} 
in the internal shock, 
where $\gamma_{\mu,\rm sup}=\epsilon_{\mu,\rm sup}/(m_\mu c^2)$
is the Lorentz factor of a muon with the energy of $\epsilon_{\mu,\rm sup}$.
Therefore, the energy of the synchrotron photons is much higher than that of the thermal photons from the jet head, 
leading to the production of $\lesssim$ GeV neutrinos via the $\Delta$-resonance of photo-meson interaction, which is beyond our scope.

On the other hand,
a low luminosity and a large Lorentz factor result in a slow acceleration according to Equation (\ref{t_acc}),
but a fast photo-meson cooling according to Equations (\ref{t_Delta}) and (\ref{tpg}).
For the extreme case $t_{\Delta}\leq t_{\rm p, acc}(\epsilon_{\rm p,th})$, 
the protons are quickly cooled via the photo-meson interaction as long as they are accelerated to be above the threshold energy.
Then the spectrum is no longer an extended power law spectrum, 
but a peaked spectrum with peak energy around $\epsilon_{\rm p,th}$.

\begin{figure}
\includegraphics[width=95mm,angle=0]{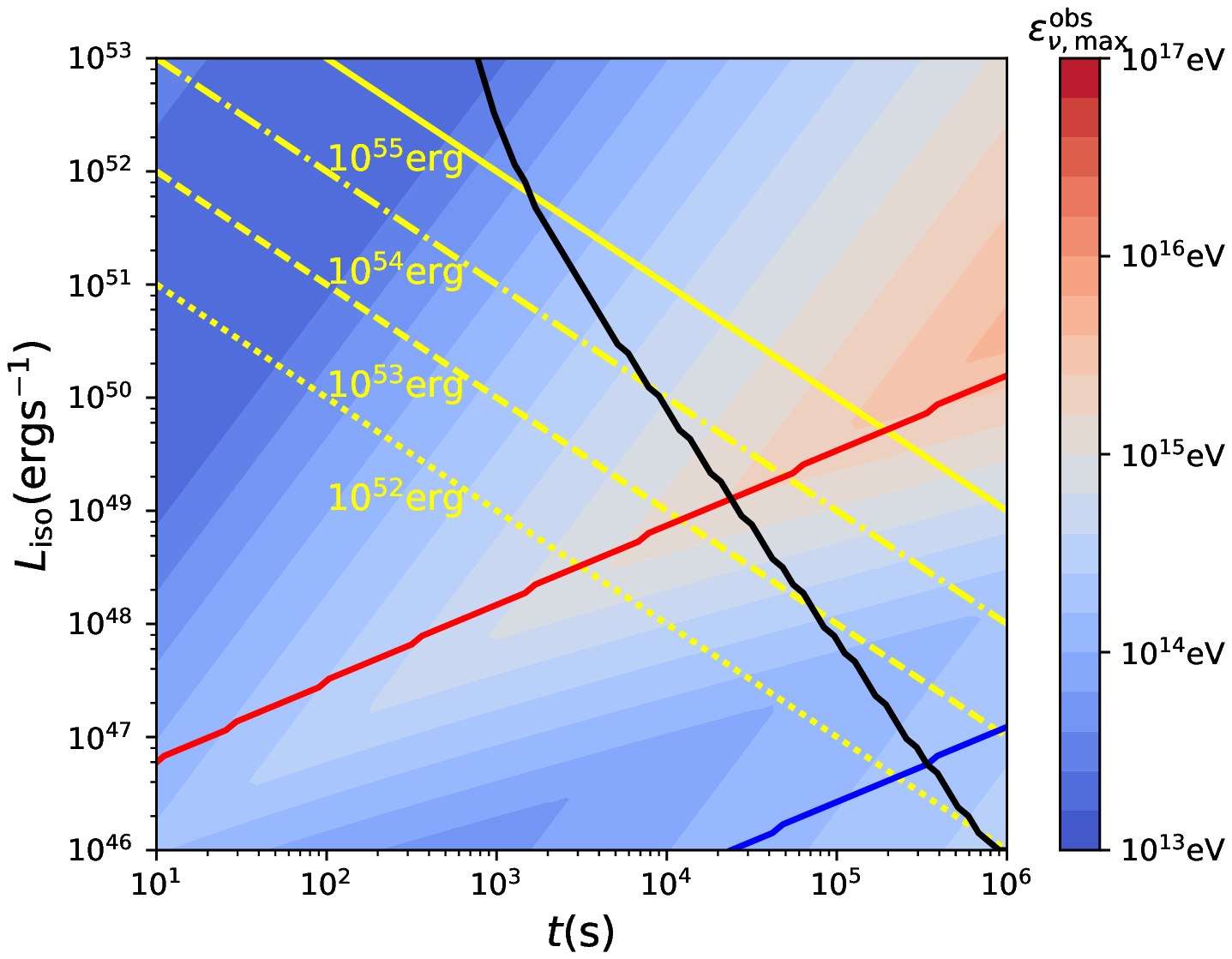}
\includegraphics[width=95mm,angle=0]{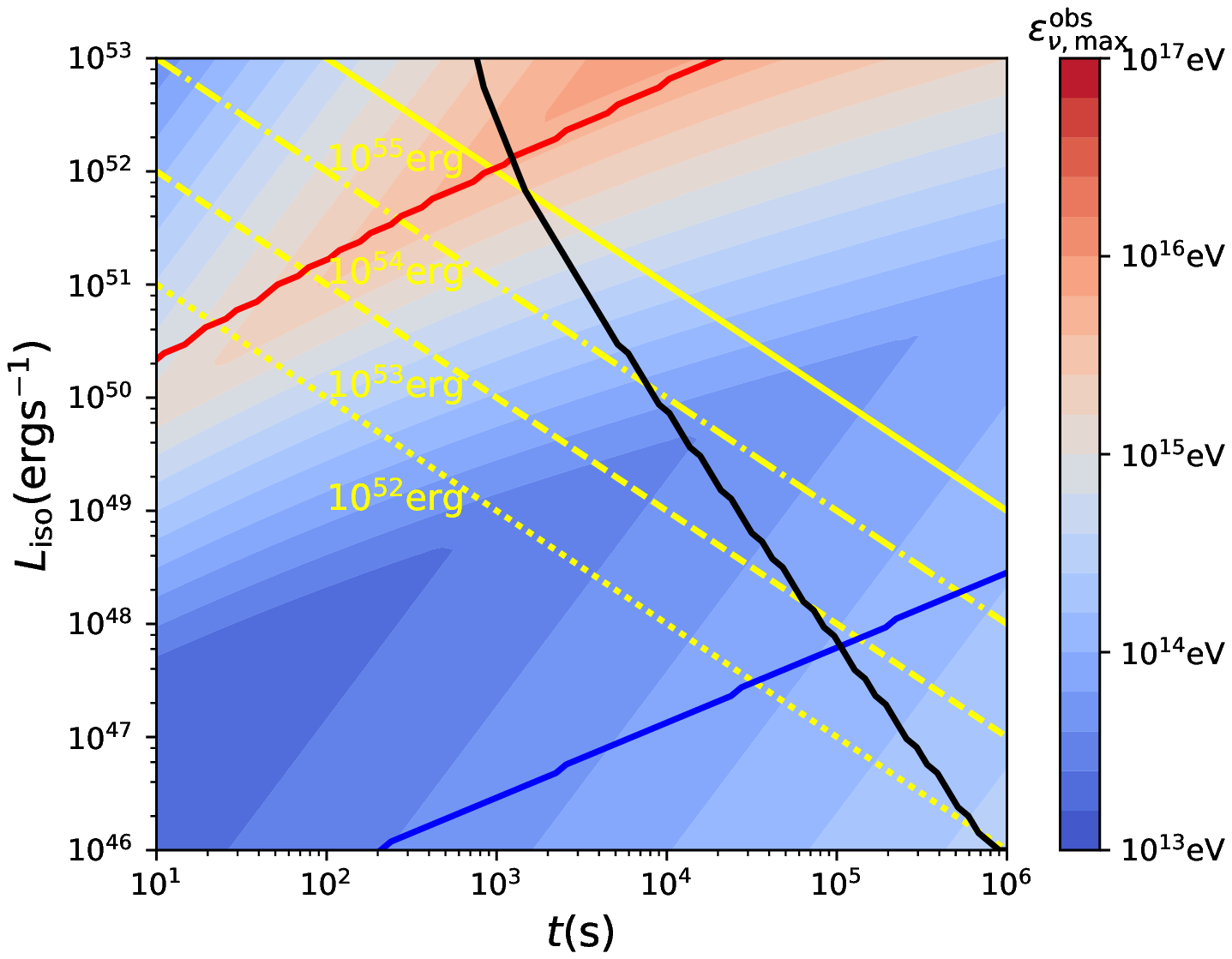}
 \caption{
Color contours denoting the estimated maximum energy of the observed neutrinos $\epsilon_{\nu,\rm max}^{\rm obs}$, 
for Lorentz factor $\Gamma=10$ (upper panel) and $\Gamma=100$ (bottom panel),
respectively.
The solid lines denote the same constraints as in Figure \ref{epsilon_th}.
}
\label{E_cut}
\end{figure}

We estimate the overall maximum energy of the produced neutrinos by comparing 
the three energies calculated above,  
which are
$\epsilon_{\nu,\rm c}^{\rm obs}$, $\epsilon_{\nu,\rm sup}^{\rm obs}$, and $\epsilon_{\nu,\rm th}^{\rm obs}$. 
The overall maximum energy of the produced neutrinos is
\begin{equation}\label{Enu_max}
\epsilon_{\nu,\rm max}^{\rm obs}\simeq\max(\epsilon_{\nu,\rm th}^{\rm obs},\min(\epsilon_{\nu,\rm c}^{\rm obs},\epsilon_{\nu,\rm sup}^{\rm obs})).
\end{equation}
The fast cooling case where $\epsilon_{\nu,\rm max}^{\rm obs}=\epsilon_{\nu,\rm th}^{\rm obs}$ can be realized
for a larger Lorentz factor and lower luminosity.
We plot the value of $\epsilon_{\nu,\rm max}^{\rm obs}$ in Figure \ref{E_cut}.

From Figure \ref{E_cut}, one can see that the parameter space is roughly divided into two regions:
the red region corresponds to a maximum energy of the spectrum $\geq 10^{15}$ eV, which is hereafter called the ``hard phase",
while the dark blue region corresponds to a maximum energy of the spectrum $\leq 10^{14}$ eV, which is called the ``soft phase",
and the light blue region corresponds to a maximum energy of the spectrum in between $10^{14}$ and $10^{15}$ eV, which is called the ``intermediate phase".

As shown in the upper panel in Figure \ref{E_cut},
for a low Lorentz factor, $\Gamma=10$,
it is unlikely that the spectrum is hard enough to explain PeV neutrinos; this is 
due to the photo-meson cooling.
On the other hand, 
as in the bottom panel in Fig \ref{E_cut},
for the case with a large Lorentz factor, $\Gamma=100$,
a high luminosity of the jet is required to explain PeV neutrinos.

In the next section,
we adopt a few parameter sets for different phases with the same isotropic injection energy $E_{\rm iso}=L_{\rm iso}t=10^{53}~{\rm erg}$,
to study the neutrino spectra from individual sources for different phases.

\section{neutrinos from individual sources}

To calculate the neutrino spectrum distribution, we adopt the analytical description  
of the energy distribution of neutrinos 
produced in the photo-meson interaction from \citet{Kelner2008}.
We also consider the effect of photo-meson cooling on the proton acceleration and 
the suppressions due to the synchrotron cooling of the secondary pions and muons, 
as discussed in Section \ref{TheHighEnergyCutoff}.

\begin{figure}
\includegraphics[width=95mm,angle=0]{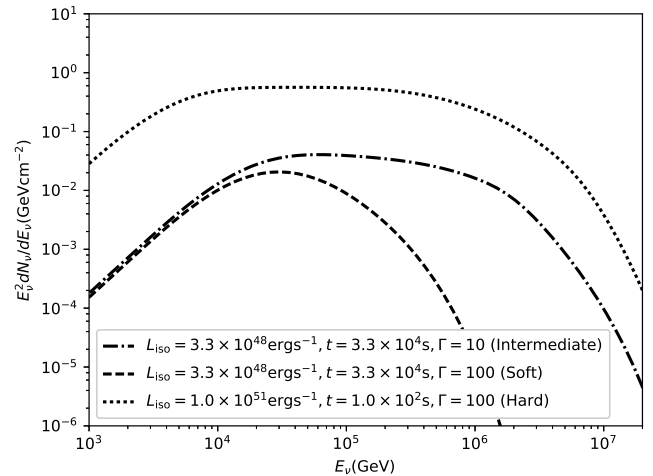}
 \caption{
Fluence of neutrinos produced from individual sources at a distance of 1 Gpc for three parameter sets.
The other parameters are fixed as $\epsilon_{\rm B}=0.01$, $\epsilon_{\rm e}=0.1$, $p=2.0$, $R_{\rm H}=3\times10^{13}~{\rm cm}$, 
and $\rho_{\rm H}=10^{-7}~{\rm g~cm^{-3}}$.
The dashed, dash--dotted, and dotted lines denote the soft, intermediate, and hard phases, respectively. 
}
\label{spectrum1Gpc}
\end{figure}

According to Figure \ref{E_cut}, 
we choose one parameter set of $L_{\rm iso}=3.3\times10^{48}~{\rm erg ~s^{-1}}$ and $t=3.3\times10^{4}~{\rm s}$,
which is in the soft phase for $\Gamma=100$,
while is in the intermediate phase for $\Gamma=10$.
We also choose the parameter set of $L_{\rm iso}=1.0\times10^{51}~{\rm erg ~s^{-1}}$ and $t=1.0\times10^2~{\rm s}$,
which is in the hard phase for $\Gamma=100$,
while is not valid for $\Gamma=10$ since it violates the acceleration constraint.
The spectra of the neutrino fluence from individual sources at a distance of 1 Gpc are plotted 
in Figure \ref{spectrum1Gpc}.
We can see three different types of spectra.
The dashed line shows a soft spectrum,
which can only contribute to the $10-100$ TeV neutrinos due to the fast photo-meson cooling,
the dotted line shows a hard spectrum, extending from a few TeV to PeV energy,
and the dash--dotted line shows an intermediate phase between the soft and the hard spectrum.

\section{Diffuse neutrinos}\label{Diffuse}
To calculate the diffuse neutrino spectrum,
we assume that the rate of choked jets $R_{\rm cj}(z)$ at redshift $z$ follows the star formation rate $\rho_{\rm sf}(z)$,
i.e.,
\begin{equation}\label{Rcj}
R_{\rm cj}(z)=A_{\rm cj}\rho_{\rm sf}(z),
\end{equation}
where the star formation rate is ~\citep{Madau2014SFR}
\begin{equation}
\rho_{\rm sf}(z)=0.015~{M_\odot\rm yr^{-1}Mpc^{-3}}~\frac{(1+z)^{2.7}}{1+[(1+z)/2.9]^{5.6}},
\end{equation}
and $A_{\rm cj}$ is a free normalization coefficient in units of ${M_\odot}^{-1}$.

Here we adopt a constant luminosity instead of a luminosity function for the choked jets,
because in our model only sources with luminosities satisfying those constraints as in Figure 1
can produce neutrinos, and their luminosity distribution would not follow a known luminosity function.
The diffuse neutrino flux is calculated via integrating the neutrino spectrum over the redshift from 0 to 8, i.e.,
~\citep{Murase2007} 
\begin{equation}
\begin{split}
\epsilon_{\nu}^{{\rm obs}}\frac{dN_{\nu}}{d\epsilon_{\nu}^{\rm obs}}(\epsilon_{\nu}^{\rm obs})d\epsilon_{\nu}^{{\rm obs}}&=\frac{c}{4\pi H_0}\int_0^8\epsilon_{\nu}\frac{dN_{\nu}}{d\epsilon_{\nu}}((1+z)\epsilon_{\nu})d\epsilon_{\nu}\\
&\times\frac{\frac{\Omega}{4\pi}A_{\rm cj}\rho_{\rm sf}(z)dz}{(1+z)\sqrt{\Omega_{\Lambda}+\Omega_{\rm M}(1+z)^3}}\\
\end{split}
\end{equation}
where the cosmological parameters are adopted as
$H_0=70~{\rm km s^{-1} Mpc^{-1}}$,$\Omega_{\rm M}=0.3$, and $\Omega_{\lambda}=0.7$,
$\epsilon_{\nu}\frac{dN_{\nu}}{d\epsilon_{\nu}}(\epsilon_{\nu})d\epsilon_{\nu}$ is the neutrino flux at the source frame,
and $\Omega=2\pi(1-\cos\theta)$ is the solid angle of a jet
with the open angle assumed to be $\theta=0.2$.

\begin{table*}
\begin{center}
\begin{tabular}{cccccccccc}
\hline
&$L_{\rm iso}$&$t$&$\Gamma$& &$A_{\rm cj}$	 &$R_{\rm cj}(z=0)$&$N_{\rm S}(N_{\nu_\mu}>1)$&$N_{\rm S}(N_{\nu_\mu}>2)$&$N_{\rm S}(N_{\nu_\mu}>3)$\\
&${\rm erg s^{-1}}$&s&  	&&$M_\odot^{-1}$&${\rm Gpc^{-3}yr^{-1}}$&${\rm yr^{-1}}$&${\rm yr^{-1}}$&${\rm yr^{-1}}$\\
\hline
\hline
Soft Phase&$3.3\times10^{48}$&$3.3\times10^{4}$&100&  &$1.4\times10^{-3}$	&$2.1\times10^4$&2.0&0.77&0.42\\
Intermediate Phase&$3.3\times10^{48}$&$3.3\times10^{4}$&10&  &$3.0\times10^{-4}$	&$4.5\times10^3$&2.1&0.78&0.42\\
Hard Phase&$1.0\times10^{51}$&$1.0\times10^2$&100&  &$1.0\times10^{-4}$	&$1.5\times10^3$&2.5&0.81&0.45\\
\hline
\end{tabular}
\end{center}
\caption{
Isotropic luminosity $L_{\rm iso}$, 
Lifetime $t$, and Lorentz Factor $\Gamma$ of the Jet, Constrained Normalization Parameter $A_{\rm cj}$,
and Constrained Local Rate of the Choked jets $R_{\rm cj}$ for the Single-component Fitting in Figure 4.
In the last three columns, $N_{\rm S}(N_{\nu_\mu}>1)$, $N_{\rm S}(N_{\nu_\mu}>2)$, and $N_{\rm S}(N_{\nu_\mu}>3)$
denote the expected amount of sources from which more than 1, 2, and 3 muon neutrinos can be detected by IceCube per year, respectively.
}
\label{tab3}
\end{table*}

We calculated the diffuse neutrino spectra for three parameter sets the same as in Figure \ref{spectrum1Gpc}.
The diffuse neutrinos are dominated by the contribution from sources at redshifts of around $z=1-3$,
since the star formation rate density peaks around those redshifts.
However, there are still many uncertainties 
on the total star formation rate density at high redshifts due to problems with 
selections of high-redshift galaxies and accurate measurements of their star formation rates ~\citep[e.g., ][]{Wang2016Cluster}.
It is possible that there are more massive stars at high redshift, for example, PoP III stars,
which may contribute more to the neutrinos. 

\begin{figure}
\includegraphics[width=95mm,angle=0]{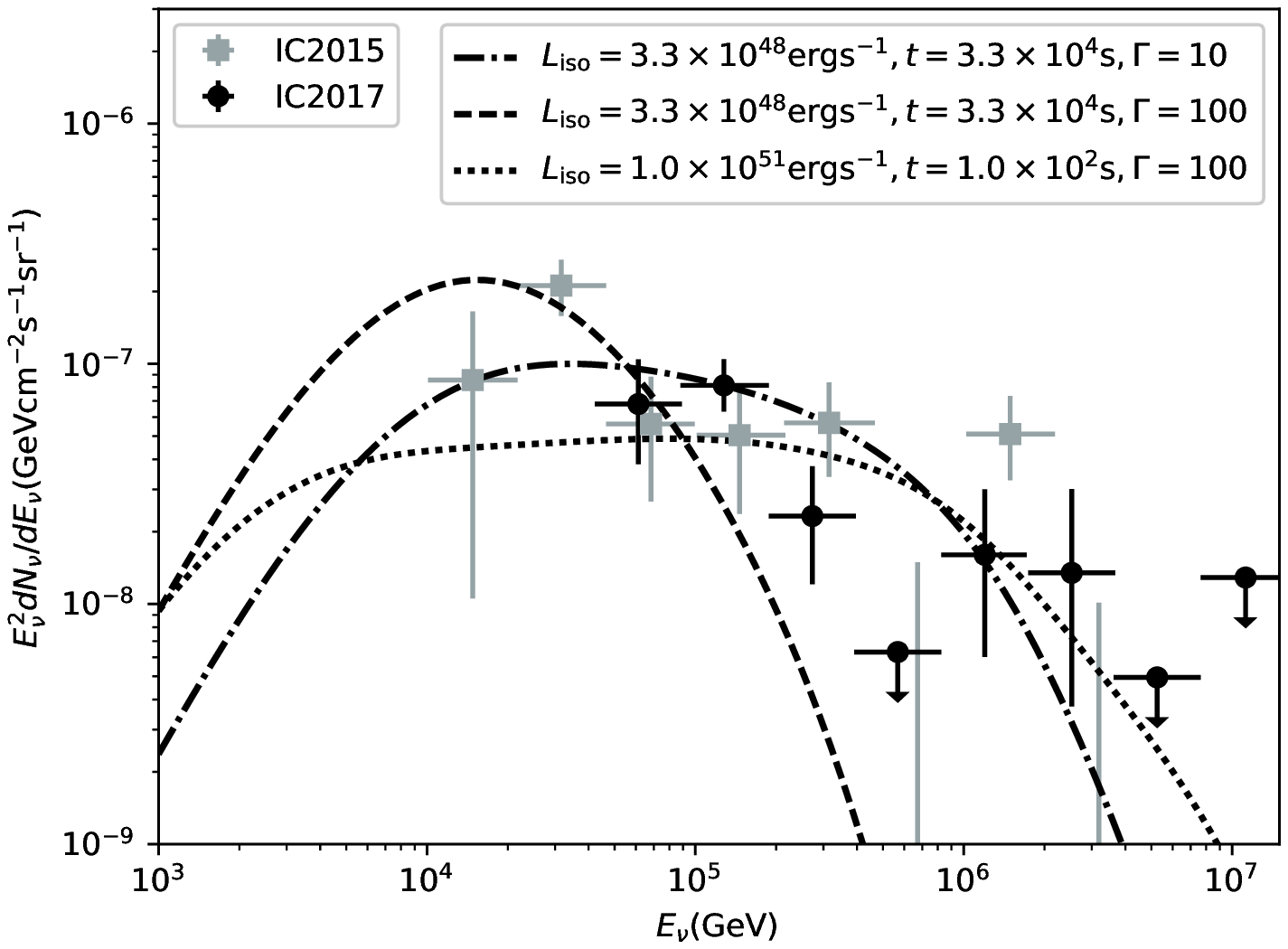}
 \caption{
Diffuse neutrino spectra for three parameter sets as in Figure \ref{spectrum1Gpc}. 
The light gray squares denote the data from the IceCube combined analysis ~\citep[IC2015,][]{IceCube2015Combined},
and the black diamonds denote the IceCube six year HESE data~\citep[IC2017,][]{IceCube2017ICRCII}.
}
\label{DiffuseSpectrum}
\end{figure}

We calculate the neutrino spectra with three fixed parameter sets, 
implying three fixed spectral shapes.
The normalization parameter $A_{\rm cj}$ is the only free parameter;
then the flux of the spectra in Figure \ref{DiffuseSpectrum} are normalized via the parameter $A_{\rm cj}$. 
By fitting the IceCube combined data ~\citep{IceCube2015Combined} and the IceCube six year HESE data~\citep{IceCube2017ICRCII} for all flavor neutrinos,
we get the best values of $A_{\rm cj}$, which are listed in Table \ref{tab3}.
The three spectra, compared with the combined 
and six year data for all flavor neutrinos,
are listed in Figure \ref{DiffuseSpectrum}.
Then the corresponding constrained local rate of the choked jet events $R_{\rm cj}(z=0)$ can be calculated via Equation (\ref{Rcj}),
and the values are also listed in Table 1.
The jet's kinetic energy, $1.0\times10^{51}~{\rm erg}$, which we adopt in this paper, is typical for a SN,
and the constrained local choked jet rate is about 
$1.5\times 10^{3}~-~2.1\times10^{4}~{\rm Gpc^{-3}~yr^{-1}}$, 
as seen in Table 1.
For comparisons, the required event rate is larger than the rate of observed successful GRBs,
where the rate of high-luminosity long GRBs (with isotropic luminosity
above $10^{50}~{\rm erg~s^{-1}}$) is $0.8^{+0.1}_{-0.1}~{\rm Gpc^{-3}~yr^{-1}}$, 
and the rate of low-luminosity long GRBs (with isotropic luminosity above $5\times 10^{46}~{\rm erg ~s^{-1}}$)  is $164^{+98}_{-65}~{\rm Gpc^{-3}~yr^{-1}}$\citep{Sun2015}.
However, the required event rate is only about $1\%-20\%$ of the typical type-II SN rate, $\sim~10^{5}~\rm Gpc^{-3}~yr^{-1}$~\citep{Atteia2013}, which is consistent with our assumption.

Since the angular resolution of observing muon neutrinos is better than that of observing the other flavors,
considering the beam correction,
we calculate the rates of observing more than 1, 2, and 3 muon neutrinos for the three parameter sets 
based on the constrained source rate for the single-component fitting, 
and list them in the last three columns of Table 1. 
Here we assume the ratio between the three flavors 
is approximated to $1:1:1$ due to their oscillations~\citep{Learned1995,Athar2000}, 
although the real flavor ratio is slightly different from $1:1:1$ considering their oscillations in the stellar envelope~\citep{Sahu2010}.
As we can see from Table 1, IceCube can observe about four triplets of muon neutrinos during 10 years of operation.

\begin{figure}
\includegraphics[width=95mm,angle=0]{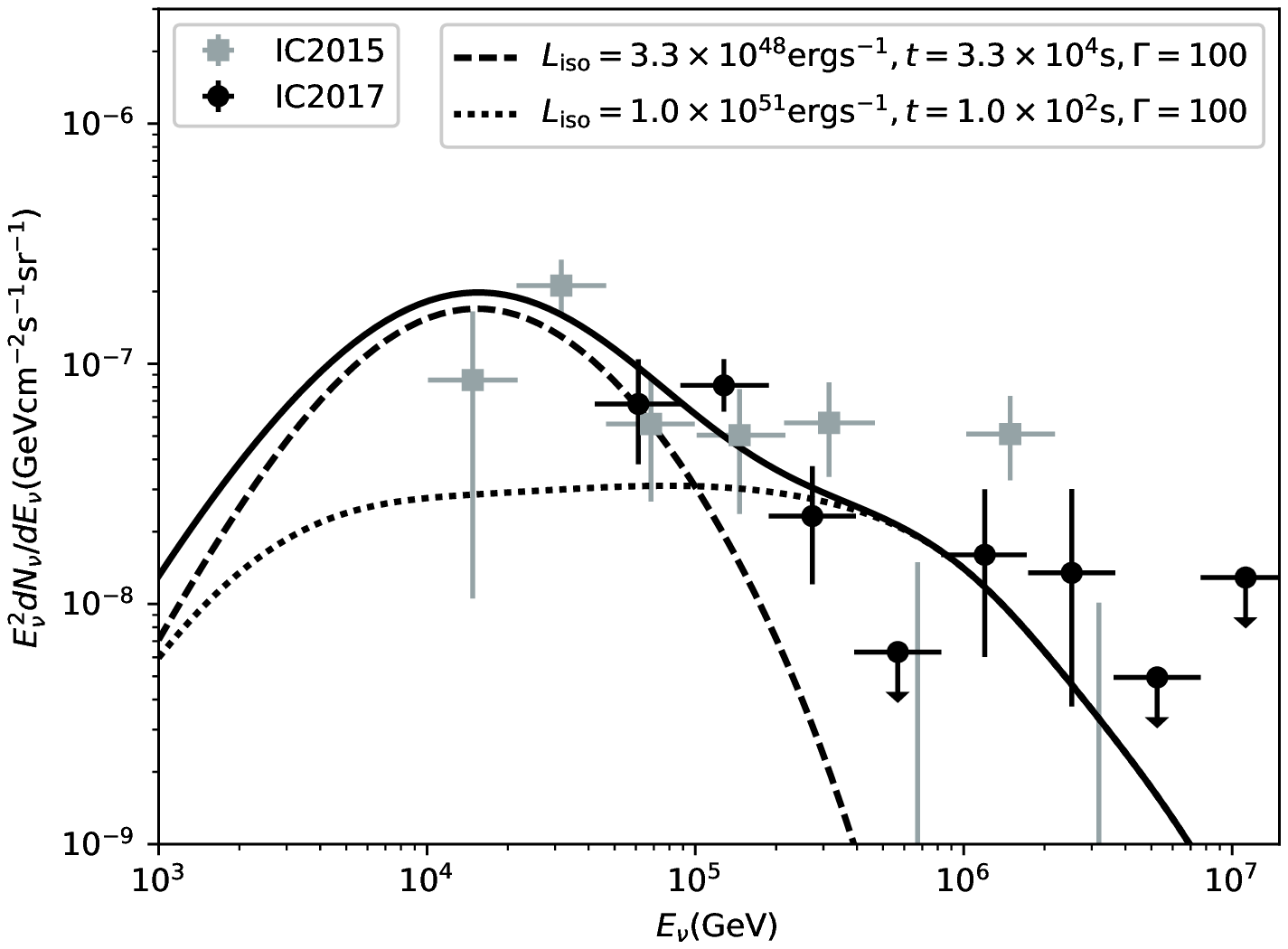}
\includegraphics[width=95mm,angle=0]{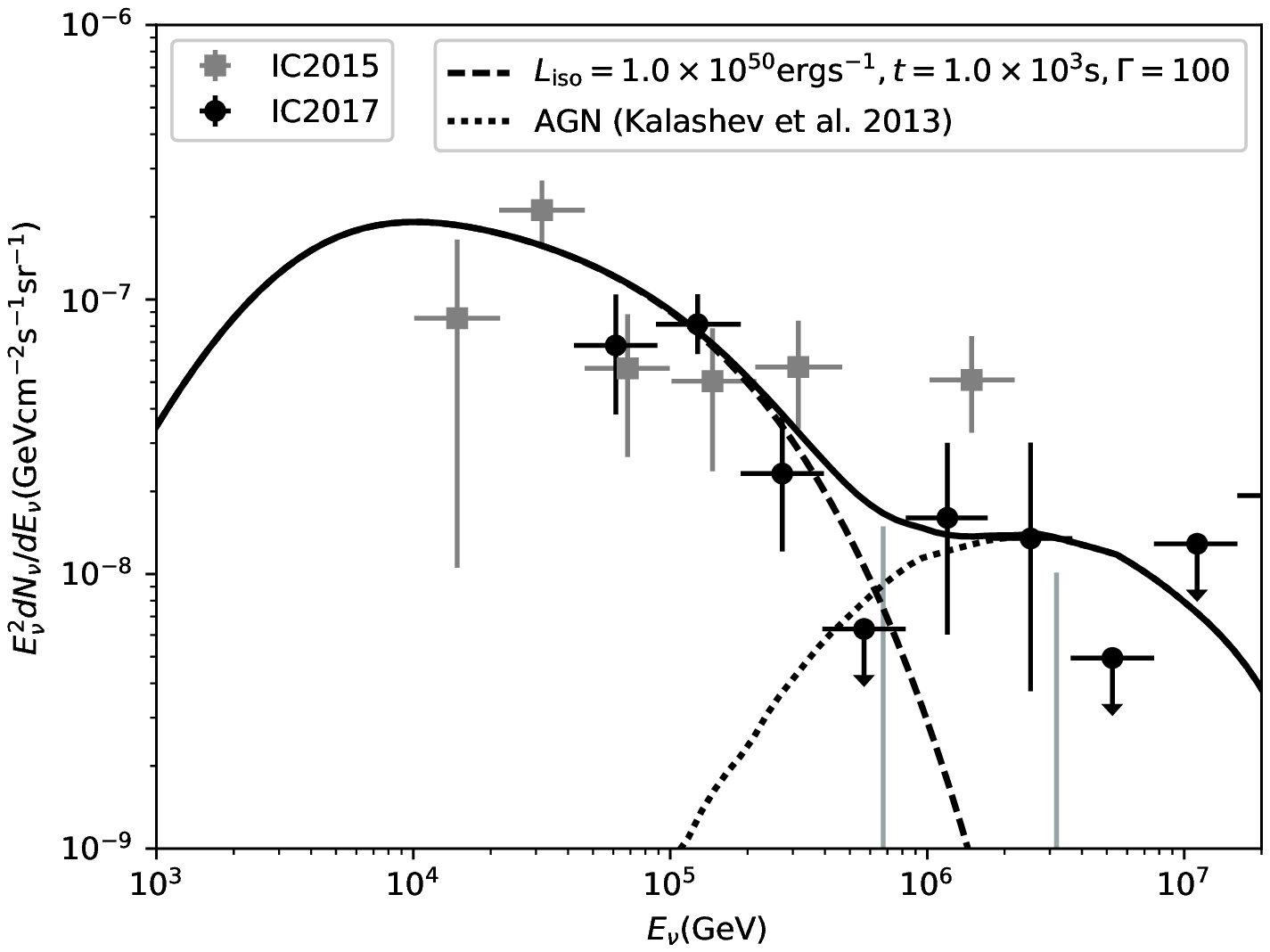}
\caption{Two-component spectra. In the upper panel, a soft phase with the parameter set of $L_{\rm iso}=3.3\times10^{48}~{\rm erg s^{-1}}$,
$t=3.3\times10^{4}~{\rm s}$, and $\Gamma=100$ is adopted as the dashed line,
and a hard phase with the parameter set of $L_{\rm iso}=1.0\times10^{51}{\rm erg s^{-1}}$, $t=1.0\times10^{2}{~\rm s}$, and $\Gamma=100$ is adopted as the dotted line.
In the bottom panel, an intermediate phase with the parameter set of $L_{\rm iso}=1.0\times 10^{50}~{\rm erg ~s^{-1}}$,
 $t=1.0\times10^{3}~{\rm s}$, and $\Gamma=100$ is adopted as the dashed line,
and a PeV neutrino component from distant blazars~\citep{Kalashev2013PRL} is adopted for the dotted line.
The other parameters are the same as in 
Figure~\ref{DiffuseSpectrum}.
The solid line is the sum of the dashed line and the dotted line.
 }
\label{2componentspectrum}
\end{figure}

As seen in Figure \ref{DiffuseSpectrum}, the hard phase (dotted line) produces a hard spectrum with a cutoff at $\sim$ PeV,
while the soft phase (dashed line) produces a soft spectrum with a cutoff at lower energy,
but cannot produce PeV neutrinos.
If we assume that both the hard and the soft phases contribute to the neutrino spectrum,
one can get a soft spectrum with a cutoff at $\sim$ PeV.
The two-component spectra, assuming that the flux ratio of the hard phase 
to the soft phase at 10 TeV is 0.2,
are plotted in the upper panel of Figure \ref{2componentspectrum}.
The existence of the cutoff at $\sim$ PeV is evidence of this model.
If the energy of the cutoff is above a few to 10 PeV, 
one needs an additional component to explain the $\geq \rm PeV$ neutrinos. 
Neutrinos from AGN cores~\citep{Stecker2005} or from distant blazars\citep{Kalashev2013PRL} may contribute at the $\geq \rm PeV$ energy,
and neutrinos from the choked jets contribute at the lower energy.
This possibility is plotted in the bottom panel of Figure \ref{2componentspectrum}.

\section{Observational predictions}\label{typeIISN}
The choked jet in our model might not result in GRBs, 
rather, it might result in a type-II~\citep[especially type-IIP,][]{Arcavi2017} SN,
since the total jet energy exceeds the gravitational binding energy of the progenitor star.
This is consistent with the observations that only a small fraction of SNe are associated with GRBs.

After the central engine stops, 
the jet tail catches up with the jet head after $\delta t_1=R_{\rm h}/\beta c$,
where $\beta c\simeq c$ is the velocity of the jet tail.
Then the jet head starts to decelerate and propagate like a jet-driven SN explosion~\citep{Nagataki2000}. 
Since the typical radial velocity of an SN shock is about $v_{\rm sh}\sim 10^9~{\rm cm~s^{-1}}$ \citep{Wongwathanarat2015},
the time for the jet head to break out through the outer stellar envelope, with length of $(R-R_{\rm h})$, is evaluated as
$\delta t_{2}\simeq (R-R_{\rm h})/v_{\rm sh}$, where $R$ and $R_{\rm h}$ are the radii of the star and the jet head.
Meanwhile, it takes $\delta t_3=(R-R_{\rm h})/c$ for neutrinos to reach the stellar surface.
Therefore, 
the observed time delay of photons from the beginning of the SN explosion to neutrinos is
$\delta t= \delta t_{1}+\delta t_{2}-\delta t_{3}=R(1/v_{\rm sh}-1/c)-R_{\rm h}(1/v_{\rm sh}-2/c)=3.2\times10^4~{\rm s}~R_{13.5}-8.4\times10^3~{\rm s}~L_{\rm iso,48}^{1/4}t_4^{1/2}\rho_{{\rm H},-7}^{-1/4}$.
We note here, since the light curve of the SN explosion takes a few days to a few tens of days to reach the peak flux, 
we may not be able to detect photons at the very beginning of the SN explosion. 
Fortunately, the light curve of Type II SNe usually extends from around a hundred days to a few hundred days,
we have a large time window to observe them.
Once we observe an SN spatially associated with a muon neutrino triplet,
we can trace back to the explosion time according to the observed light curve,
and then measure the time difference between the neutrino burst and the explosion.

As listed in Table 1, about four triplets of muon neutrinos are expected to be observed during 10 years of IceCube's observations.
The detections on new-born type-II SNe following the observation of muon neutrino singlets, doublets, 
and especially triplets, will be strong evidence for the dominant contribution of the choked jet neutrino.
Moreover, the jet-induced feature, i.e., the substantially asymmetric and predominantly bi-polar explosion\citep{Wheeler2002}, will provide us with more evidence for the existence of the choked jet.
Any future observed time difference between the muon neutrino triplet and the photon emission will help one to constrain the model.

We calculate
the mean number of observed muon neutrino events from a source at distance of $D$ via
$N_{\nu_\mu}(D)=\int F(E_{\nu_\mu},D)A_{\rm eff}(E_{\nu_\mu})dE_{\nu_\mu}$,
where $F(E_{\nu_\mu},D)$ is the fluence of neutrinos produced from an individual source at distance of $D$, 
and $A_{\rm eff}(E_{\nu_\mu})$ is the effective area of IceCube for muon neutrinos.
Then we can derive the mean upper limit of the distance of sources
from which a triplet of muon neutrinos can be observed by IceCube, 
via the criterion of $N_{\nu_\mu}(D)\geq 3$. 
The mean upper limit of the source distance is constrained to be 81 Mpc, 135 Mpc, and 188 Mpc,
for the soft, intermediate and hard phases shown in Table 1, respectively,
which are within the current detection radius of core-collapse SNe~\citep{Taylor2014}.

For an extreme high isotropic energy $E_{\rm iso}=10^{54}~{\rm erg}$, corresponding to a jet energy of $10^{52}~{\rm erg}$ for $\theta=0.2$, 
as seen in the bottom panel of Figure 2, 
only a small parameter space is consistent with our model, 
and the associated SN might be a type II superluminous SN (SLSN).
If one adopts parameters $\Gamma=100$, $L_{\rm iso}=10^{51}~{\rm erg~s^{-1}}$, and $t=10^3~{\rm s}$,
more than three muon neutrinos on average can be observed by IceCube 
if the source is located within $\sim 0.6~\rm Gpc$. 
This limitation on the source distance ($z\leq 0.05$) is within the current detection radius 
of an SLSN.

The new-born type-II SN/SLSN associated with the triplet of muon neutrinos
might be observed by optical/infrared and X-ray observatories.
The three instruments adopted in the IceCube Optical Follow-up program 
and X-ray Follow-up program~\citep{Kowalski2007, Abbasi2012, IceCube2015TypeIInSN}, namely 
the Robotic Optical Transient Search Experiment (ROTSE;~\citet{Akerlof2003}), the Palomar Transient Factory (PTF;~\citet{Law2009,Rau2009}),
and the {\it Swift} satellite~\citep{Gehrels2004}, 
as well as the Subaru Hyper-Suprime-Cam with a field of view (FOV) of $1.5^\circ$ 
will be suitable telescopes for the follow-up observations, 
since the median angular resolution of the muon track events is around $1^\circ$~\citep{IceCube2016TA}.
The Large Synoptic Survey Telescope (LSST) with a $3.5^\circ$ FOV ~\citep{Angeli2014LSST}
as well as the Pan-STARRS1 (PS1) telescope with a $3.3^\circ$ FOV,
will provide a useful archive to search for detections retrospectively~\citep{IceCube2015TypeIInSN}.

\section{Conclusions}\label{discussions}
Choked jet sources produce neutrinos without accompanying gamma rays, 
thus avoiding a conflict between the diffuse gamma-ray observations and neutrino observations.
The choked jets might be very common at the end of the massive stars' lives. 
In this paper, we have considered the properties of the neutrino spectrum from a choked jet in a red supergiant star,
which is associated with type II SNe/SLSNe.
The parameter space consistent with observations is roughly divided into the soft and hard phases.
The soft phase corresponds to a soft part of the spectrum with a cutoff at around a few tens of TeV. This component does not contribute to PeV neutrinos.
The hard phase corresponds to a hard spectrum with a cutoff at $\sim$ PeV. 
For the choked jet model, it is difficult to explain the soft feature of the spectrum and PeV neutrinos using a single component.
However, a two-component spectrum can explain the IceCube data. 
In addition to a soft spectrum contributing to the low-energy neutrinos,  
a second component, such as a hard spectrum, neutrinos from AGN cores~\citep{Stecker2005} 
or neutrinos from distant blazars~\citep{Kalashev2013PRL} contributing to the PeV neutrinos, is needed. 
We plot one-component spectra and two-component spectra in Figures \ref{DiffuseSpectrum}, and
\ref{2componentspectrum}, respectively.
The contribution from a fast photo-meson cooling case softens the neutrino spectrum,
which is consistent with the currently observed soft spectrum. 
Another predicted feature is the low-energy cutoff at $\sim 20$ TeV, as discussed in Section 5.1.
A possible cutoff around ~PeV can be used to test the origin of the second component.

Neutrino oscillations can affect the observed flavor composition depending on whether the neutrinos go through a stellar envelope~\citep{Sahu2010}. 
Hence, the future precise observations on the neutrino flavor ratio can be another way to test the choked jet model.

Furthermore, our model predicts newly born jet-driven type-II SNe/SLSNe associated with the production of neutrinos.
The detection of such SNe associated with an observation of a muon neutrino singlet,
doublet, or triplet will be a strong evidence in favor of the choked jet model.

\acknowledgements
We thank Ruo-Yu Liu, Hirotaki Ito, Ji-An Jiang, Kohta Murase and Peter M$\rm\acute{e}$sz$\rm\acute{a}$ros for the useful comments.
This work is supported in part by the Mitsubishi Foundation,
a RIKEN pioneering project Interdisciplinary Theoretical Science (iTHES) 
and Interdisciplinary Theoretical $\&$ Mathematical Science Program (iTHEMS).
H.N.H. is supported by National Natural Science of China under grant 11303098, and
the Special Postdoctoral Researchers (SPDR) Program in RIKEN.
The work of A.K. is supported by the U.S. Department of Energy Grant No. DE-SC0009937 and by the World Premier International Research Center Initiative (WPI), MEXT, Japan.
S.N. is supported by JSPS (Japan Society for the Promotion of Science):
No.25610056, 26287056, MEXT (Ministry of Education, Culture, Sports, Science and Technology): No.26105521,
and Mitsubishi Foundation in FY2017.
Y.Z.F. is supported by 973 Program of China under grant 2013CB837000.
D.M.W. and Y.Z.F. are supported by 973 Programme of China (No. 2014CB845800), by NSFC under grants 11525313 (the National Natural Fund for Distinguished Young Scholars), 11273063, 11433009 and 11763003.
\bibliographystyle{apj}

\end{document}